\documentclass[12pt]{iopart}

\usepackage{iopams}
\usepackage{graphicx}
\usepackage{dcolumn}
\usepackage{bm}
\usepackage{float}
\newcommand{\bra}[1]{\langle#1|}

\newcommand{\ket}[1]{|#1\rangle}

\usepackage{color}
\newcommand{\blue}[1]{{\color{blue}#1}}

\begin{document}

\title{Dynamics of correlations due to a phase noisy laser}

\author{Bruno Bellomo$^1$, Rosario Lo Franco$^{1,2}$, Erika Andersson$^3$, James D. Cresser$^4$, and Giuseppe Compagno$^1$}
\address{$^1$CNISM and Dipartimento di Fisica, Universit\`a di Palermo, via Archirafi 36, 90123 Palermo, Italy\\
$^2$Centro Siciliano di Fisica Nucleare e Struttura della Materia (CSFNSM) and Dipartimento di Fisica e Astronomia, Universit\`a di Catania, Viale A. Doria 6, 95125 Catania, Italy\\
$^3$SUPA, Department of Physics, Heriot-Watt University, Edinburgh EH14 4AS, United Kingdom\\
$^4$Centre for Quantum Computer Technology, Department of Physics and Engineering, Macquarie University, 2109 NSW, Australia}

\begin{abstract}
We analyze the dynamics of various kinds of correlations present between two initially entangled independent qubits, each one subject to a local phase noisy laser. We give explicit expressions of the relevant quantifiers of correlations for the general case of single-qubit unital evolution, which includes the case of a phase noisy laser. Although the light field is treated as classical, we find that this model can describe revivals of quantum correlations. Two different dynamical regimes of decay of correlations occur, a Markovian one (exponential decay) and a non-Markovian one (oscillatory decay with revivals) depending on the values of system parameters. In particular, in the non-Markovian regime, quantum correlations quantified by quantum discord show an oscillatory decay faster than that of classical correlations. Moreover, there are time regions where nonzero discord is present while entanglement is zero.
\end{abstract}


\maketitle

\section{\label{introduction}Introduction}
Total correlations present in a system can be separated in a purely quantum part and a classical part. Quantum correlations can be classified as those associated with non-separability (entanglement) and other quantum correlations, together quantified by the quantum discord (QD) \cite{Zurek2001PRL,Henderson2001JPA}. For quantum information processing it is relevant to know the dynamics of these different kinds of correlations.
Entanglement dynamics can be considered well understood, in its general lines, for bipartite quantum systems interacting with quantum environments (independent or common), presenting phenomena like sudden death \cite{yu2004PRL}, revivals \cite{bellomo2007PRL,bellomo2008PRA,mazzola2009PRA} or trapping \cite{bellomo2008trapping}, depending on the Markovian or non-Markovian nature of the environments. Dynamics of QD has also been investigated for two-qubit systems in the presence of both Markovian \cite{maziero2010PRA,ferraro2010PRA} and non-Markovian \cite{fanchini2010PRA,wang2010PRA} quantum environments.

In this paper we analyze the dynamics of correlations in a two-qubit system where each qubit is subject to a phase noisy laser modeled as a classical field. This is a commonly used model for the interaction between light and matter, and thus the dynamics of correlations for this situation merits investigation. The broader question underlying our work is whether all kinds of behavior of correlations that can be described using a quantum environment also can be well described using a classical approximation of the environment. One might for example expect that a classical environment should not be able to store quantum correlations on its own, and that therefore revivals of quantum correlations in the system may be affected. For the phase-noisy laser, are there qualitative features that will be lost by treating the field classically as opposed to quantum-mechanically? Among other things, we find that this model can describe both decay and oscillatory behavior of quantum correlations.

\section{Model: a qubit subject to a phase-noisy laser}
Our system consists of a pair of qubits (two-level atoms), $A$ and $B$, each driven by a local phase noisy laser. Each atom under the action of its own laser is described, in a rotating frame and for resonant atom-field interaction, by the Hamiltonian \cite{Cresser2009}
\begin{equation}\label{Hamiltonian}
    \hat{H}=\lambda \left[\sigma_-\mathrm{e}^{\mathrm{i}\Phi(t)}+\sigma_+\mathrm{e}^{-\mathrm{i}\Phi(t)}\right],
\end{equation}
where the laser is described as a classical field with a randomly fluctuating phase $\Phi(t)$. The interaction between each qubit and its local field mode is assumed to be strong enough so that, for sufficiently long times, the dissipation effects of the vacuum radiation modes on the qubit dynamics can be neglected. This could be feasible by considering, as a qubit, an atom in a cavity subject to a resonant interaction with the phase noisy laser but out of resonance with cavity mode frequencies in order to inhibit effects like spontaneous emission. In this phase-noisy model the phase undergoes a Wiener process, i.e. $\Phi(t)$ is white noise with a correlation function $\langle\dot{\Phi}(t)\dot{\Phi}(t+\tau)\rangle=2d\delta(\tau)$ where $d$ is a diffusion rate. In Eq.~(\ref{Hamiltonian}), $\lambda$ is the atom-field coupling strength, $\sigma_+=\ket{1}\bra{0}$ and $\sigma_-=\ket{0}\bra{1}$ are the atomic raising and lowering operators, where $\ket{0}$ and $\ket{1}$ are the ground and excited state of the atom, respectively. The field correlation function, corresponding to the above phase correlation function, is a complex colored noise $e^{i\Phi(t)}$ described by $\langle\mathrm{e}^{i \Phi}(t)\mathrm{e}^{-i \Phi}(t+\tau)\rangle=\mathrm{e}^{-d\tau}$.

The Hamiltonian of Eq.~(\ref{Hamiltonian}) leads to a local-in-time non-Markovian master equation of the form \cite{Cresser2009}
\begin{equation}\label{unitalmastereq}
\dot{\rho}(t)=\sum_{i=1}^3(\gamma_i\sigma_i\rho\sigma_i-\gamma_i\rho),
\end{equation}
where $\sigma_i$ ($i=1,2,3$) are the atomic pseudo-spin Pauli matrices and
\begin{equation}\label{gammaandGamma}
\gamma_1=\gamma_2=-\frac{\dot{\Gamma}_1}{4\Gamma_1},\ \gamma_3=-\frac{1}{2}\left(\frac{\dot{\Gamma}_2}{\Gamma_2}
-\frac{\dot{\Gamma}_1}{2\Gamma_1}\right).
\end{equation}
This master equation is derived by the Nakajima-Zwanzig projection operator method and the functions $\Gamma_1,\Gamma_2$ can be expressed in terms of the system parameters. In particular $\Gamma_1(t)=\mathrm{e}^{-\frac{dt}{2}}\left[\cosh\left(\frac{1}{2}t\sqrt{d^2-16\lambda^2}\right)
+d\sinh\left(\frac{1}{2}t\sqrt{d^2-16\lambda^2}\right)/\sqrt{d^2-16 \lambda ^2}\right]$, while the expression for $\Gamma_2(t)$ is cumbersome and is not given here. Because the $\Gamma_i$ are time-dependent, the master equation of Eqs.~(\ref{unitalmastereq}) and (\ref{gammaandGamma}) has a quasi-Lindblad form, that is a master equation that resembles the Lindblad form but has time-dependent decay rates $\gamma_i$ (which also can be negative) \cite{Cresser2009}. Moreover, the above single-qubit master equation is a unital master equation \cite{andersson2007JMO}. A master equation $\partial\rho/\partial t=\mathcal{L}_t(\rho)$ is defined to be \emph{unital} if the maximally mixed state $\frac{1}{2}\hat{I}$ is a fixed point, that is if $\mathcal{L}_t(\hat{I})\equiv0$. Unital master equations can be solved by Kraus-type decomposition methods giving the single-atom reduced density matrix elements at time $t$ as \cite{andersson2007JMO}
\begin{eqnarray}\label{singleatomreduceddensitymatrix}
\rho_{11}(t)&=&[(1+\Lambda_3)\rho_{11}(0)+(1-\Lambda_3)\rho_{00}(0)]/2,\nonumber\\
\rho_{10}(t)&=&[(\Lambda_1+\Lambda_2)\rho_{10}(0)+(\Lambda_1-\Lambda_2)\rho_{01}(0)]/2,
\end{eqnarray}
with $\rho_{11}=1-\rho_{00}$, $\rho_{10}=\rho^*_{01}$. All the $\Lambda_i(t)\equiv\Lambda_i$ are time-dependent and related to the rates $\gamma_i$ by
\begin{equation}\label{Lambdaandgamma}
\Lambda_i(t)=\mathrm{e}^{-2 \int_0^t \mathrm{d}t'[\gamma_j(t')+\gamma_k(t')]},
\end{equation}
with the conditions $\Lambda_i+\Lambda_j\leq1+\Lambda_k$, where $\{i,j,k\}$ run over the cyclic permutations of $\{1,2,3\}$. Note also that $\Lambda_j(0)=1$ and $\Lambda_j(t)\geq0$. Using Eq.~(\ref{gammaandGamma}) in Eq.~(\ref{Lambdaandgamma}) we have $\Lambda_1=\Lambda_2=\Gamma_2$, $\Lambda_3=\Gamma_1$ and the resulting reduced density matrix elements for the single atom driven by a phase noisy laser reduces, from Eq.~(\ref{singleatomreduceddensitymatrix}), to $\rho_{11}(t)=\frac{1}{2}[(1+\Gamma_1)\rho_{11}(0)+(1-\Gamma_1)\rho_{00}(0)]$ and $\rho_{10}=2\Gamma_2\rho_{10}(0)$.
It has been shown that for $d/\lambda<4$, $\Gamma_1$ oscillates while $\Gamma_2$ shows oscillatory behavior only for $d/\lambda <2.606$ \cite{Cresser2009}; for $d/\lambda>4$, neither $\Gamma_1$ nor $\Gamma_2$ shows oscillatory behavior. Finally, in the limit $d/\lambda\rightarrow0$, the two decay rates $\Gamma_1,\Gamma_2$ become periodic functions of $\lambda t$, more specifically $\Gamma_1\rightarrow\cos(2\lambda t)$ and $\Gamma_2\rightarrow\cos^2(\lambda t)$. In this limit, the probability distribution of the phase is a steady state distribution, so that the effective dynamics is the statistical average, equally weighted, of all the evolutions with phase between $0$ and $2\pi$. An analogous but simpler setup, described by a uniform phase distribution with only two values ($0,\pi$), has been investigated in Ref.~\cite{LoFranco2010arxive}.

\section{Quantifiers of two-qubit correlations\label{sec:quantifiers of correlations}}
To obtain the expressions of the correlation quantifiers we need the two-qubit density matrix elements. We construct the two-qubit reduced density matrix at time $t$ by the knowledge of the evolution of single-qubit reduced density matrices, according to a standard procedure \cite{bellomo2007PRL}, with the single-qubit reduced density matrix evolution given by Eqs.~(\ref{singleatomreduceddensitymatrix}) and (\ref{Lambdaandgamma}). Thus, we obtain the explicit expressions of the two-qubit density matrix elements at time $t$ in terms of the functions $\Lambda_i$ for any initial two-qubit state. These expressions are however quite cumbersome and are not reported here.

We take as two-qubit initial states the extended Werner-like (EWL) states \cite{bellomo2008PRA}
\begin{equation}\label{EWLstates}
    \hat{\rho}^\Phi=r \ket{\Phi}\bra{\Phi}+\frac{1-r}{4}I_4,\quad
    \hat{\rho}^\Psi=r \ket{\Psi}\bra{\Psi}+\frac{1-r}{4}I_4,
\end{equation}
where $r$ indicates the purity of the initial states, $I_4$ is the $4\times4$ identity matrix,
$\ket{\Phi}=\alpha\ket{01}+\beta\mathrm{e}^{i\delta}\ket{10}$ and $\ket{\Psi}=\alpha\ket{00}+\beta\mathrm{e}^{i\delta}\ket{11}$ are the Bell-like states where $\alpha,\beta$ \blue{are} non-negative real numbers and $\alpha^2+\beta^2=1$. These are mixed states reducing to Werner states for $\alpha=\beta=\pm1/\sqrt{2}$ or to Bell-like states for $r=1$. The density matrix elements of the EWL states are such that the resulting density matrix has an ``X'' structure with nonzero elements only along the main diagonal and anti-diagonal.

In the standard basis $\mathcal{B}=\{\ket{1}\equiv\ket{11},\ket{2}\equiv\ket{10},\ket{3}\equiv\ket{01},\ket{4}\equiv\ket{00}\}$ and for general different environments characterized by different values of $\Lambda_i^S$ $(i=1,2,3;\ S=A,B)$, the time-dependent two-qubit density matrix elements for $\hat{\rho}^\Phi$ are
\begin{eqnarray}\label{twoqubitelements}
&\rho^\Phi_{jj}(t)=\frac{1}{4}\left\{1-r\left[\Lambda^A_3\Lambda^B_3
+(-1)^j(1-2\alpha^2)(\Lambda^A_3-\Lambda^B_3)\right]\right\},&\nonumber\\
&\rho^\Phi_{ll}(t)=\frac{1}{4}\left\{1+r\left[\Lambda^A_3\Lambda^B_3
+(-1)^l(1-2\alpha^2)(\Lambda^A_3+\Lambda^B_3)\right]\right\},&\nonumber\\
&\rho^\Phi_{1+k4-k}(t)=\frac{\alpha\beta r}{2}\left[f(\Lambda)\cos\delta
+\mathrm{i}(-1)^{k+1}g(\Lambda)\sin\delta\right],&
\end{eqnarray}
where $j=1,4$, $l=2,3$, $k=0,1$, $f(\Lambda)=\Lambda^A_1\Lambda^B_1+\Lambda^A_2\Lambda^B_2$ and $g(\Lambda)=\Lambda^A_1\Lambda^B_2+\Lambda^A_2\Lambda^B_1$. The density matrix elements for the initial state $\hat{\rho}^\Psi$ are obtained from Eq.~(\ref{twoqubitelements}) by changing $1\leftrightarrow2$ and $3\leftrightarrow4$. Note that under the dynamical conditions here considered, the two-qubit state maintains an X structure during the evolution.

In order to describe the entanglement dynamics we use the concurrence, which for an X state is given by \cite{yu2007QIC}
$C_\rho^X(t)=2\mathrm{max}\{0,K_1(t),K_2(t)\}$ where $K_1(t)=|\rho_{23}(t)|-\sqrt{\rho_{11}(t)\rho_{44}(t)}$ and $K_2(t)=|\rho_{14}(t)|-\sqrt{\rho_{22}(t)\rho_{33}(t)}$. The EWL states of Eq.~(\ref{EWLstates}) present the same initial value of the concurrences, $C(0)=2\mathrm{max}\{0,(\alpha\beta+1/4)r-1/4\}$, from which one finds that there is initial entanglement when $r>r^\ast=(1+4\alpha\beta)^{-1}$. Using the time-dependent density matrix elements of Eq.~(\ref{twoqubitelements}), it is readily seen that in our system the concurrence at time $t$ is the same for both the EWL states of Eq.~(\ref{EWLstates}), that is $C_\rho^\Phi(t)=C_\rho^\Psi(t)=C(t)$. For example, for initial Bell states ($r=1, \alpha=1/\sqrt{2},\delta=0,\pi$) and different local conditions, concurrence is given by $C(t)=\frac{1}{2}\left(\Lambda^A_1 \Lambda^B_1+\Lambda^A_2\Lambda^B_2+\Lambda^A_3 \Lambda^B_3-1\right)$.

In order to quantify total correlations, $\mathcal{T}$, present in the two-qubit system and to distinguish a quantum $\mathcal{D}$ and a classical part $\mathcal{J}$ of them, we use the notion of quantum discord \cite{Zurek2001PRL,Henderson2001JPA}. The calculation of discord and classical correlations requires a maximization procedure and this has been analytically solved only for certain class of quantum states (Bell-diagonal and X states) \cite{Luo2008PRA,ali2010PRA}. In particular, here we give the explicit expressions of these quantifiers for the initial condition $\alpha=\beta=1/\sqrt{2}, \delta=0,\pi$ for which the two-qubit density matrix always has a Bell-diagonal form $\rho_\mathrm{B}=[\mathbb{I}\otimes\mathbb{I}+\sum_{j=1}^3c_{j}\sigma_j\otimes\sigma_j]/4$. For this class of states, the following expressions for $\mathcal{T}$ and $\mathcal{J}$, 
with $\mathcal{D}=\mathcal{T}-\mathcal{J}$, hold \cite{Luo2008PRA}:
\begin{eqnarray}\label{total, discord and classical}
   \mathcal{T}=2+\sum_{i,s}\lambda_i^s\mathrm{log}\lambda_i^s \quad
    \mathcal{J}=\sum_{i}^2 \frac{1+(-1)^ic}{2}\mathrm{log}[1+(-1)^ic],
\end{eqnarray}
where $i=1,2; s=\pm$, $c\equiv\mathrm{max}\{|c_1|,|c_2|,|c_3|\}$, $\lambda_1^\pm=(1\pm c_1\pm c_2-c_3)/4$, $\lambda_2^\pm=(1\pm c_1\mp c_2+c_3)/4$. If the initial state is $\hat{\rho}^\Phi$, the $c_i$ coefficients are $c_1=r\mathrm{max}\{\Lambda^A_1 \Lambda^B_1,\Lambda^A_2 \Lambda^B_2 \}$, $c_2=r\mathrm{min}\{\Lambda^A_1\Lambda^B_1,\Lambda^A_2\Lambda^B_2 \}$ and $c_3=-r\Lambda^A_3\Lambda^B_3$. If the initial state is $\hat{\rho}^\Psi$, previous coefficients change as $c_1\rightarrow c_1$, $c_2\rightarrow -c_2$ and $c_3\rightarrow-c_3$, that is a relabeling of the $\lambda$ eigenvalues: the quantifiers $\mathcal{T}$, $\mathcal{D}$ and $\mathcal{J}$ thus coincide for both initial states.

\section{Dynamics of correlations due to a phase-noisy laser}
\begin{figure}
\begin{center}
{\includegraphics[width=0.45 \textwidth]{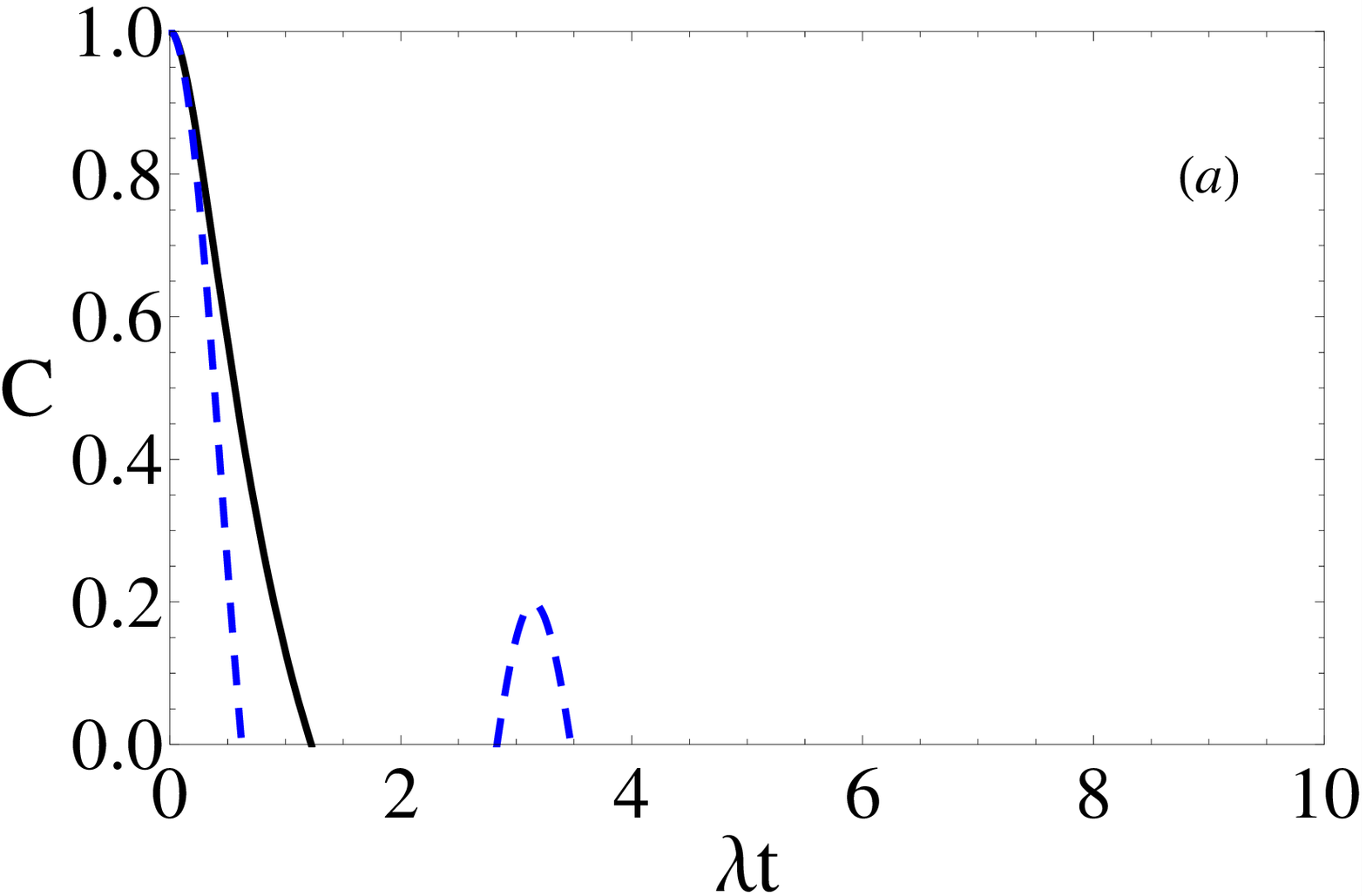}
\hspace{0.5 cm}
\includegraphics[width=0.45 \textwidth]{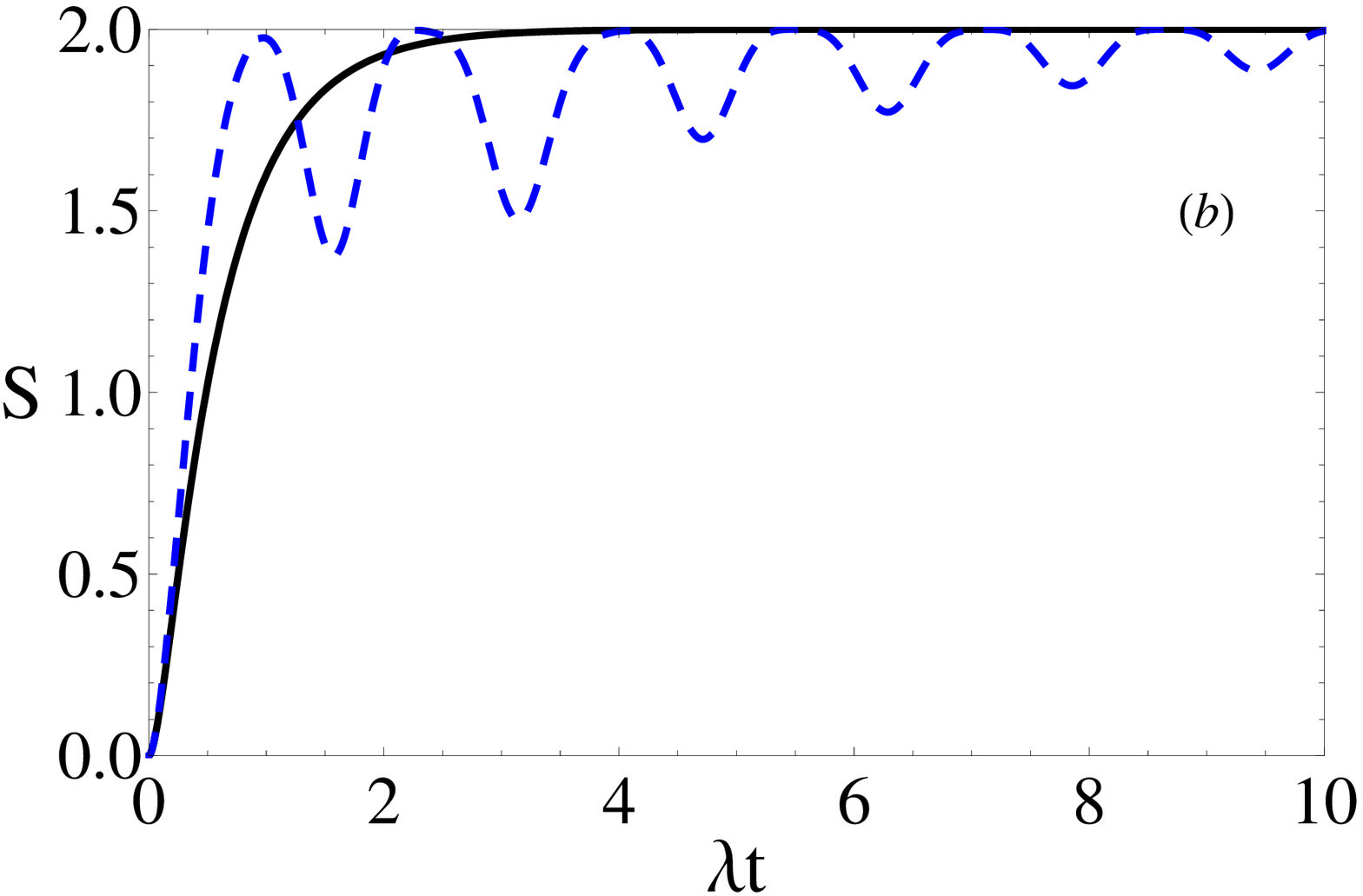}}
\caption{\label{Concurrence and entropy}\footnotesize Time evolution of concurrence $C$ (panel (a)) and von Neumann entropy $S$ (panel (b)) starting from initial Bell states ($r=1,\alpha=1/\sqrt{2},\delta=0, \pi$) for the ratios $d/\lambda=5$ (black solid curve) and $d/\lambda=0.1 $ (blue dashed curve).}
\end{center}
\end{figure}

We use the general results of the previous section to analyze the dynamics of correlations. In particular, for initial EWL states in identical environments ($\Gamma_j^A=\Gamma_j^B=\Gamma_j,\, j=1,2$) we obtain $C_\rho^\Phi(t)=C_\rho^\Psi(t)=C=\frac{1}{2}\left(4 r \alpha\beta\Gamma_2^2+r \Gamma^2_1-1\right)$. In Fig.~\ref{Concurrence and entropy}(a) we plot concurrence evolution for initial Bell states, for two different values of the ratio $d/\lambda=5,0.1$. The behavior for more general initial mixed states can be shown to be qualitatively similar. One sees that while concurrence presents Markovian-like decay for $d/\lambda= 5$, it is subject to non-Markovian revival for $d/\lambda=0.1$, as already found in other systems \cite{bellomo2007PRL,LoFranco2010arxive}. In order to evidence the fact that for large times the two-qubit system goes toward a maximally mixed state as a consequence of the unital evolution of the single qubits, in Fig.~\ref{Concurrence and entropy}(b) we plot the evolution of the von Neumann entropy, $S=-\mathrm{Tr}\{\rho(t)\log\rho(t)\}$, for the same values of $d/\lambda$ (von Neumann entropy is the same for both initial EWL states $S_\rho^\Phi(t)=S_\rho^\Psi(t)=S$). It is seen that, for any value of $d/\lambda$, $S$ tends to its maximum value 2, corresponding to the two-qubit maximally mixed state, with oscillating behavior when memory effects are present ($d/\lambda=0.1$).

\begin{figure}
\begin{center}
\includegraphics[width=0.45 \textwidth]{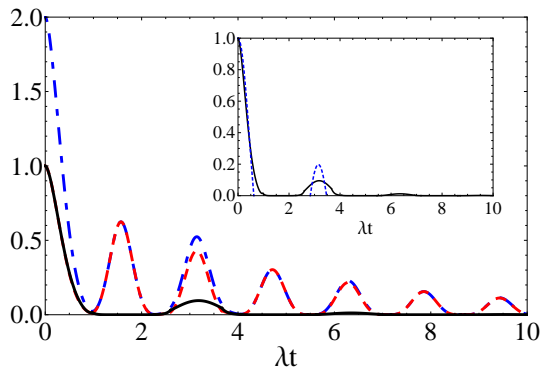}
\caption{\label{quantum vs classical}\footnotesize Time evolution of quantum discord $\mathcal{D}$ (black solid curve), classical correlations $\mathcal{J}$ (red dashed curve) and total correlations $\mathcal{T}$ (blue dot-dashed curve), starting from initial Bell states ($r=1, \alpha=1/\sqrt{2}, \delta=0, \pi$) for the $d/\lambda=0.1$. In the inset $\mathcal{D}$ (black solid curve) and concurrence $C$ (blue dashed curve) are plotted.}
\end{center}
\end{figure}

We can also separately investigate the evolution of quantum and classical correlations. In Fig.~\ref{quantum vs classical} we plot the evolution of total, quantum and classical correlations of Eq.~(\ref{total, discord and classical}) for the value of the ratio $d/\lambda=0.1$ for initial Bell states.
From the plot one sees that, in this non-Markovian case, total and classical correlations show an oscillating decay while quantum discord has a non-purely oscillating decay (i.e., there are points of discontinuity for the derivative during the time regions where discord is very close to zero) and it decays much faster than classical correlations. On the other hand, as seen from the inset of Fig.~\ref{quantum vs classical}, discord definitively disappears later than entanglement, with time regions when it is different from zero while entanglement is absent.

\section{Conclusions \label{par:Conclusion}}
In this paper we have analyzed the dynamics of correlations between two initially entangled independent qubits each locally subject to a phase noisy laser. We found explicit expressions for various quantifiers of correlations, such as concurrence and quantum discord, in terms of the decay rates present in the single-qubit unital master equation which here describes the action of phase noisy laser. We discussed how the dynamics of correlations depends on the ratio between the phase diffusion rate $d$ and the atom-laser coupling strength $\lambda$.

Although the light field is treated as classical, this model can describe revivals of quantum correlations.
For large values of $d/\lambda$ Markovian-like decay occurs, while non-Markovian effects become relevant for small values of this ratio, giving place for example to revival of entanglement and of discord. Moreover, in the non-Markovian regime, quantum discord presents an oscillatory decay faster than that of classical correlations, with time regions where it is nonzero and entanglement is zero. It is therefore evident that the phase diffusion rate and coupling strength of each atom-laser affect the qualitative time-behavior of quantum correlations in a crucial way.

\section*{References}
\providecommand{\newblock}{}

\end{document}